\documentclass[aps,prb,twocolumn,superscriptaddress]{revtex4}
\usepackage{graphicx}
\usepackage{color}

\newcommand{\be}{\begin{equation}}
\newcommand{\ee}{\end{equation}}
\newcommand{\ba}{\begin{eqnarray}}
\newcommand{\ea}{\end{eqnarray}}
\newcommand{\baa}{\begin{eqnarray*}}
\newcommand{\eaa}{\end{eqnarray*}}

\def\be{\begin{equation}}
\def\ee{\end{equation}}
\def\bea{\begin{eqnarray}}
\def\eea{\end{eqnarray}}

\def\C60{A$_x$C$_{60}$}

\def\HgCu3{HgCa$_2$Cu$_3$O$_{8+y}$}
\def\HgCu4{HgBa$_2$Ca$_3$Cu$_4$O$_{10+y}$}
\def\TlCu{Tl$_2$Ba$_2$CuO$_{6+\delta}$}
\def\TlCu3{Tl$_2$Ba$_2$Ca$_2$Cu$_3$O$_{10+y}$}
\def\TlCu4{Tl$_2$Ba$_2$Ca$_3$Cu$_4$O$_{12+y}$}

\def\BiCu3{Bi$_2$Sr$_2$Ca$_{2}$Cu$_3$O$_y$}

\def\8LSCO{La$_{1.88}$Sr$_{.12}$CuO$_4$}
\def\110LNSCO{La$_{1.5}$Nd$_{0.4}$Sr$_{0.1}$CuO$_{4}$}
\def\stage4LCO{La$_{2}$CuO$_{4+\delta}$}
\def\Y248{YBa$_2$Cu$_4$O$_8$}

\def\NbSe2{NbSe$_2$}
\def\TaSe2{TaSe$_2$}
\def\TiSe2{TiSe$_2$}

\begin{document}
\title{Longitudinal mode of spin fluctuations in iron-based superconductors}

\author{Qiang Zhang}
\affiliation{Beijing National Laboratory for Condensed Matter Physics, and Institute of Physics, Chinese Academy of Sciences, Beijing 100190, China}
\author{Jiangping Hu}\email{jphu@iphy.ac.cn }
 \affiliation{Beijing National Laboratory for Condensed Matter Physics, and Institute of Physics, Chinese Academy of Sciences, Beijing 100190, China}
\affiliation{Collaborative Innovation Center of Quantum Matter, Beijing, China}
\affiliation{Kavli Institute of Theoretical Sciences, University of Chinese Academy of Sciences, Beijing 100049, China}

\begin{abstract}
 Iron-based superconductors can exhibit different magnetic ground states and are in a critical magnetic region where frustrated magnetic interactions strongly compete with each other. Here we investigate the longitudinal modes of spin fluctuations in an unified effective magnetic model for iron-based superconductors. We focus on the collinear antiferromagnetic phase and calculate the behavior of the longitudinal mode when different phase boundaries are approached. The results can help to determine the nature of the magnetic fluctuations in iron-based superconductors. 
\end{abstract}


\maketitle

\section{\label{s:intro}Introduction}
Iron-based superconductors have very rich magnetic properties\cite{dai-np12,Dai-RMP15}. They exhibit many intriguing magnetically ordered ground states, including stripe-like collinear antiferromagnetic (CAF) state\cite{de-na08}, checkerboard-like antiferromagnetic (AFM) state\cite{Marty-PRB11}, bi-collinear antiferromagnetic (BCAF) state\cite{bao-prl09}, staggered dimmer (DI)\citep{cao-prb15} state and some incomensurate (IC) states. The superconductivity appears to be linked to the magnetism, in particular, the CAF state\cite{mazin-prl08}. The origin of these magnetic states thus has been one of central focus in this field.

 Although both itinerant and local spin theories are reasonably successful in explaining magnetic properties of some certain families of iron-based superconductors, it has been crystal clear that the magnetism is a hybrid with dual characters from both itinerant electrons and local spin moments\citep{Hu-PRB12,Glasbrenner-NP15}. However, microscopically, the system can not be simply described by a Kondo lattice type of model because it is very difficult to separate itinerant electrons from localized ones. 

A reasonable strategy is to seek an effective magnetic model. With the existence of local magnetic moments, we can still focus on the effective interactions between these local moments by integrating out of itinerant electrons to obtain a minimum magnetic effective model by keeping those Heisenberg-type leading interactions with the shortest distances. This approach has yielded a successful effective model, the $J_1-J_2-J_3-K$ model\cite{Hu-PRB12,Glasbrenner-NP15,fang-EPL09,fang-prb08,prb09-kpaper,wysocki-np11, mazin-prb08,wang-nc11}. In this model, the nearest neighbor (NN) magnetic interaction $J_1$ and the next NN one $J_2$ arise mainly through local magnetic direct exchange and magnetic superexchange mechanisms. The third NN interaction $J_3$ and the quartic interaction $K$ indicate the existence of the strong couplings to itinerant electrons. The phase diagram of the model has been studied extensively\cite{Hu-PRB12,Glasbrenner-NP15}. The mean-field results of the model can account for most magnetic phases and low energy magnetic excitations observed experimentally in iron-based superconductors\citep{Hu-PRB12} except some recently observed orders in very specific situations such as the double-$\mathbf{Q}$ orders\citep{giovannetti-nc11,jorg-prb16}. The magnetism in the effective model is extremely frustrated due to the strong competition among $J_1$, $J_2$ and $J_3$, which is also consistent with the fact that the long range magnetic order is absent in some iron-based superconductors.

In a model based on local magnetic moments, the spin waves (SW) are the low energy excitations in a given magnetically ordered state. The SW are the transverse modes, namely  the magnetic fluctuations perpendicular to the direction of the ordered moment. The longitudinal modes (LM), which are parallel to the ordered moments, are gapped out  as Higgs mode. 
However, if there are several competing magnetic states, the LM can start to appear at low energy even at zero temperature. Therefore, the gaps of the LM (Higgs mass) can provide us important information about the degree of magnetic frustration\cite{haldane-prl83,Wang-PRX13,Luo-Hu-PRL13}. 

Recently, several polarized neutron scattering experiments have been carried out in the CAF state of iron-based superconductors\cite{Wang-PRX13,qureshi-prb12,song-prb13,Luo-Hu-PRL13,dai-prb17}. The gapped LM have been observed. The observed gaps of these LM are much lower than the band width of spin excitations. Thus, these measurements suggest the existence of strong magnetic frustration in the materials. The materials may be close to a quantum critical point or are located close to a spin liquid region\cite{zaliznyak-15spin,zhou-nc13}. 

Longitudinal excitations was viewed as support for itinerant magnetism\citep{you-prb11,Eremin-prb10,Wang-PRX13}. In this paper, we use a new method to analytically calculate the LM from the effective exchange model, in particular, in the parameter region of the CAF state near the phase boundary. We find that the LM become visible at low energy close to the phase boundary and they have different dispersion relations from the transverse SW modes. They disperse very rapidly along the antiferromagnetic direction and have very little dispersion along the FM direction in the CAF phase. This feature is absent in other magnetically ordered states so that it is unique for the CAF state. Therefore, our results suggest that the measurement of the LM in a paramagnetic state that has a finite magnetic correlation length can be used to determine how the system is close to the CAF state. 

\section{\label{s:model}The $J_1-J_2-J_3-K$ Model}
\begin{figure}[ht] 
\centering 
	\includegraphics[width=.45\textwidth]{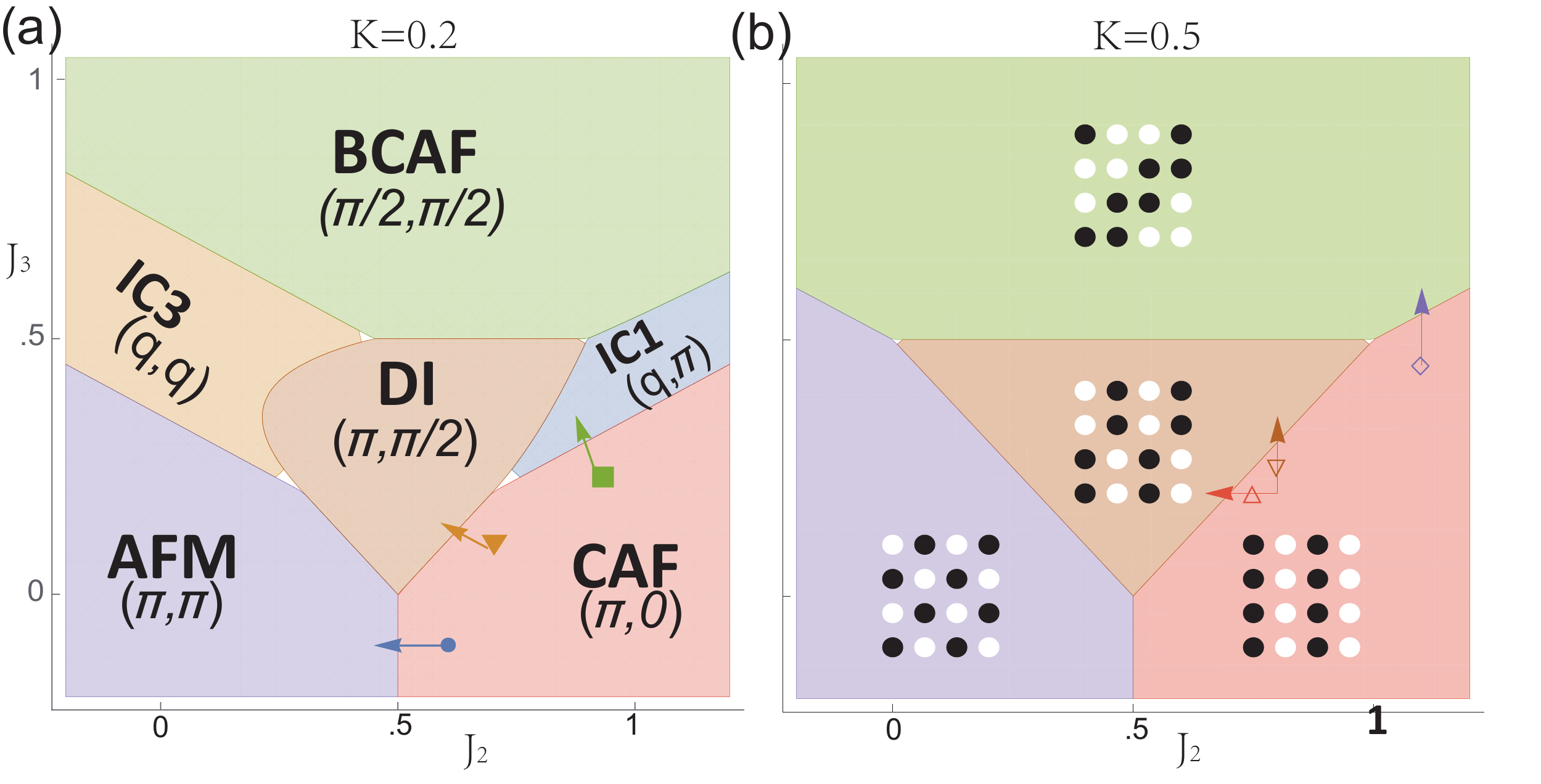}
	\caption{Various classical states for $K=0.2$ (a) and $K=0.5$ (b) in the unit of $J_1$ $(s=1)$ as classified in literatures\cite{Glasbrenner-NP15,Hu-PRB12}. For small $K$ in (a), the incommensurate phases insert in the middle of the parameter space. For bigger $K$ (b), the commensurate phases expand to cover the whole parameter space.  The  black-white dots represent the up-down spins in real space  for commensurate phases.  The  filled inverted triangle marks FeSe at $9\ GPa$\citep{Glasbrenner-NP15} and the  arrows indicate the paths approaching the phase boundaries .}\label{f:phases}
 \end{figure}

 The $J_1-J_2-J_3-K$ model\cite{Hu-PRB12,Glasbrenner-NP15} is described by
\begin{eqnarray}
H&=&\sum_{\langle i,j\rangle} J_1 \vec{S_i}\cdot\vec{S_j}+\sum_{\langle\langle i,j\rangle\rangle} J_2\vec{S_i}\cdot \vec{S_j}\nonumber\\
&& +\sum_{\langle\langle\langle i,j\rangle\rangle\rangle} J_3\vec{S_i}\cdot \vec{S_j}-\sum_{\langle i,j\rangle} K(\vec{S_i}\cdot\vec{S_j})^2,
\end{eqnarray}
which includes the nearest ($J_1$), second ($J_2$) and third ($J_3$) nearest neighbor Heisenberg interactions  and $K$($>0$) quartic term\cite{chandra-prl90,prb09-kpaper}.  Various magnetic phases can be classified in   Fig.\ref{f:phases}. We will focus on the LM in the CAF phase and the behavior of approaching the phase boundaries . \\ 
 
This Hamiltonian can be solved within the standard spin wave ({\it Holstein-Primakov}) theory. Starting from the above depicted classical ground state, we rotate the down spins and then represent the spins $\vec{S}_i$ with magnons $a_i,a^\dagger_i$:
\begin{eqnarray}
\begin{array}{ll}
S^z_i\rightarrow -S^z_i,& S^{\pm}_i\rightarrow -S^{\mp}_i,\\
S^z_i=s-a^\dagger_i a_i, & S^+_i(S^-_i)=\chi a_i( a^\dagger_i\chi),
\end{array}
\end{eqnarray}
with $\chi=\sqrt{2s-a^\dagger_ia_i}$. When the local environment for each spin is not identical, there are several spins in one magnetic cell and the site $i=\{c,a\}$ with $c=1,\cdots,N^\prime$ counting for the cells and $a=1,\cdots,n$ for the magnon types. Up to two-magnon operators (we would omit the vector notation for position and momentum), the approximate Hamiltonian in momentum space is
 \begin{eqnarray}
H\approx \sum_k\frac 12s\Psi^\dagger_kH_k\Psi_{-k}-\sum_k\frac 12s Tr(H_k)+E_0,
\end{eqnarray}
with $H_k$ the $2n\times 2n$ matrix, $E_0$ the classical ground state energy and $\Psi^\dagger_k=(\mathbf{a}^\dagger_k,\mathbf{a}^T_{-k})$, where $\mathbf{a}_k$ is a column collection of magnon annihilators and $\mathbf{a}^T_k$ is the matrix transpose. As derived in the appendix.(\ref{a:derivation}), this Hamiltonian can be diagonalized by \textit{Bogoliubov} transformation\cite{xiao-09bogoliubov} and the SW spectra $\epsilon_{nk}$ are the eigenvalues of $\sigma_{zn}H_k$. Here
\begin{equation}
H_k=\left(\begin{array}{cc}
\omega_k & \gamma_k\\ \gamma^\dagger_k & \omega^T_{-k} 
\end{array}\right),\qquad 
\sigma_{zn}=\left(\begin{array}{cc}
I_n & 0\\
0 & -I_n
\end{array}\right)
\end{equation}
with $I_n$ the $n\times n$ identical matrix and $\omega_k=\omega^\dagger_k=\omega^T_{-k}$, $\gamma_k=\gamma^T_{-k}=\gamma^\dagger_k$. When $n=1$, it is easy to find $\epsilon_k=s\sqrt{\omega_k^2-\gamma_k^2}$. This SW spectrum is generally gapless at the order momentum as the \textit{Goldstone} mode. Coupling anisotropy can interpret the observed gap\citep{Wang-PRX13}. We would shift the Heisenberg to  \textit{XXZ} coupling $\vec{S_i}\cdot\vec{S_j}=S^x_iS^x_j+S^y_iS^y_j+AS^z_iS^z_j$, with $A=1+\delta_a$. This positive $\delta_a$ will pick up the easy axis and speed up the numerical calculations.

\subsection{The Longitudinal Mode}
In quantum antiferromagnet, the longitudinal modes are the amplitude fluctuation of the ordered moments $\langle  S^z\rangle $, essentially, magnon density. Different from the itinerant approach\citep{Eremin-prb10} and nonlinear-$\sigma$ model\citep{haldane-prl83,you-prb11}, LM can also be viewed as two magnon resonance\citep{affleck-prb92} or  magnon density wave\citep{Xian-JPCS14} in Holstein-Primakov theory. Following Feynman's approach to the helium superfluid\cite{feynman-pr54}, the LM can be defined as:
\begin{eqnarray}
|L_q\rangle \equiv \frac 1{\sqrt{N}}\sum_i e^{i\vec{q}\cdot\vec{r}_i}S^z_i|0\rangle=\frac{1}{\sqrt{N}}\sum_k a^\dagger_ka_{k+q}|0\rangle.
\end{eqnarray}
With respect to the ground state, the LM has the spectrum:
\begin{eqnarray}
E(q)=\frac{N(q)}{S(q)},\label{EqSq}
\end{eqnarray}
with $N(q)\equiv \langle L_q|H|L_q\rangle$ and $S(q)\equiv \langle L_q|L_q\rangle$ the structure factor. Separating the Hamiltonian by neighborhood ($\vec{m}$) coupling: $H=\sum_mH_m+\sum_mH^K_m$, we have 
\begin{eqnarray}\label{e:commutator}
N(q)=\sum_mN_m(q)+\sum_mN^K_m(q)
\end{eqnarray} 
with (dependent on whether the $\vec{m}$ neighbor spins are anti-parallel or parallel (AP/P))
\begin{eqnarray}
N_m(q)=&\left \{
\begin{array}{rl}
\frac{J_m}{2n}\sum_a\bigl(\cos (qm)+1\bigr)\Pi_{m+},& AP\\
\frac{J_m}{2n}\sum_a\bigl(\cos (qm)-1\bigr)\Pi_{m-}, & P
\end{array}
\right.\\
N^K_m(q)=&\left \{
\begin{array}{rl}
\frac{Ks^2}{n}\sum_a \bigl(\cos (qm)+1\bigr)\Pi^K_{m+}, & AP\\
-\frac{Ks^2}{n}\sum_a\bigl(\cos (qm)-1\bigr)\Pi^K_{m-}, & P
\end{array}
\right.
\end{eqnarray}
where $\sum_a$ sums over different magnon types and the correlation function $\Pi$'s are (see appendix.\ref{a:derivation})
\begin{eqnarray}
\Pi_{m+}=&2s\Delta_{ab} -\Delta_{ab} (\rho_{aa} +\rho_{bb} )-\rho_{ab} (\Delta_{aa} +\Delta_{bb} ),\nonumber\\
\Pi_{m-}=&2s\rho_{ab} -\rho_{ab} (\rho_{aa} +\rho_{bb} )-\Delta_{ab} (\Delta_{aa} +\Delta_{bb} ),\nonumber\\
\Pi^K_{m+}=&8\Delta^2_{ab} +4\Delta_{aa} \Delta_{bb} +A\bigl(2(s-1)\rho_{ab}\nonumber\\
&-5\rho_{ab} (\rho_{aa} +\rho_{bb} )-5/2\Delta_{ab} (\Delta_{aa} +\Delta_{bb} )\bigr),\\
\Pi^K_{m-}=&8\rho^2_{ab} +4\Delta_{aa} \Delta_{bb} +A\bigl(2(s-1)\Delta_{ab}\nonumber\\& -5\Delta_{ab} (\rho_{aa} +\rho_{bb} )-5/2\rho_{ab} (\Delta_{aa} +\Delta_{bb} )\bigr).\nonumber
\end{eqnarray}
Here $\rho_{ab}(m)\equiv \langle a^\dagger_ib_{i+m}\rangle,\ \Delta_{ab}(m)\equiv \langle a_ib_{i+m}\rangle $ are the correlations of type $a,b$ magnon at site $i,i+m$. We have omitted the neighborhood $(m)$ to simplify the notation. The quartic $K$ term modifies the exchange $J$ by $\pm 2AKs^2$ at the first order of large $s$. The $n\times n$ correlation matrices in momentum space $\rho_k,\ \Delta_k$ are the Fourier transformation of $\rho(m)$, $\Delta(m)$. The structure factor is 
\begin{eqnarray}
S(q)=\frac{1}{n}Tr\Bigl[\rho(0)+\frac {1}{N^\prime}\sum_k(\rho_k\rho_{k+q}+\Delta_k\Delta_{k+q})\Bigr].
\end{eqnarray}
In the case of one type magnon, the SW spectrum and the correlators can be solved analytically:
\begin{eqnarray}
\begin{array}{l}\label{e:solve}
\epsilon_k=s\sqrt{\omega^2_k-\gamma^2_k},\\
\rho_k=\bigl(\sqrt{\omega_k^2/(\omega_k^2-\gamma_k^2)}-1\bigr)/2,\\ 
\Delta_k=-\gamma_k/(2\sqrt{\omega_k^2-\gamma_k^2}).
\end{array}
\end{eqnarray}

\subsection{\label{ss:CAF}The CAF Phase}
The CAF phase with ordered momentum $\mathbf{Q}=X=(\pi,0)$ is an example of the exactly solvable case of Eq.(\ref{e:solve}) with
\begin{eqnarray}
\omega_k=&2(J_1-2AKs^2)\cos k_y+4AJ_2+8A^2Ks^2\nonumber\\
	 	&+2J_3(\cos 2k_x+\cos 2k_y-2A),\\
\gamma_k=&-2(J_1+2AKs^2+2J_2\cos k_y)\cos k_x.\nonumber
\end{eqnarray}

\begin{figure}[ht] 
\centering 
	\includegraphics[width=.45\textwidth]{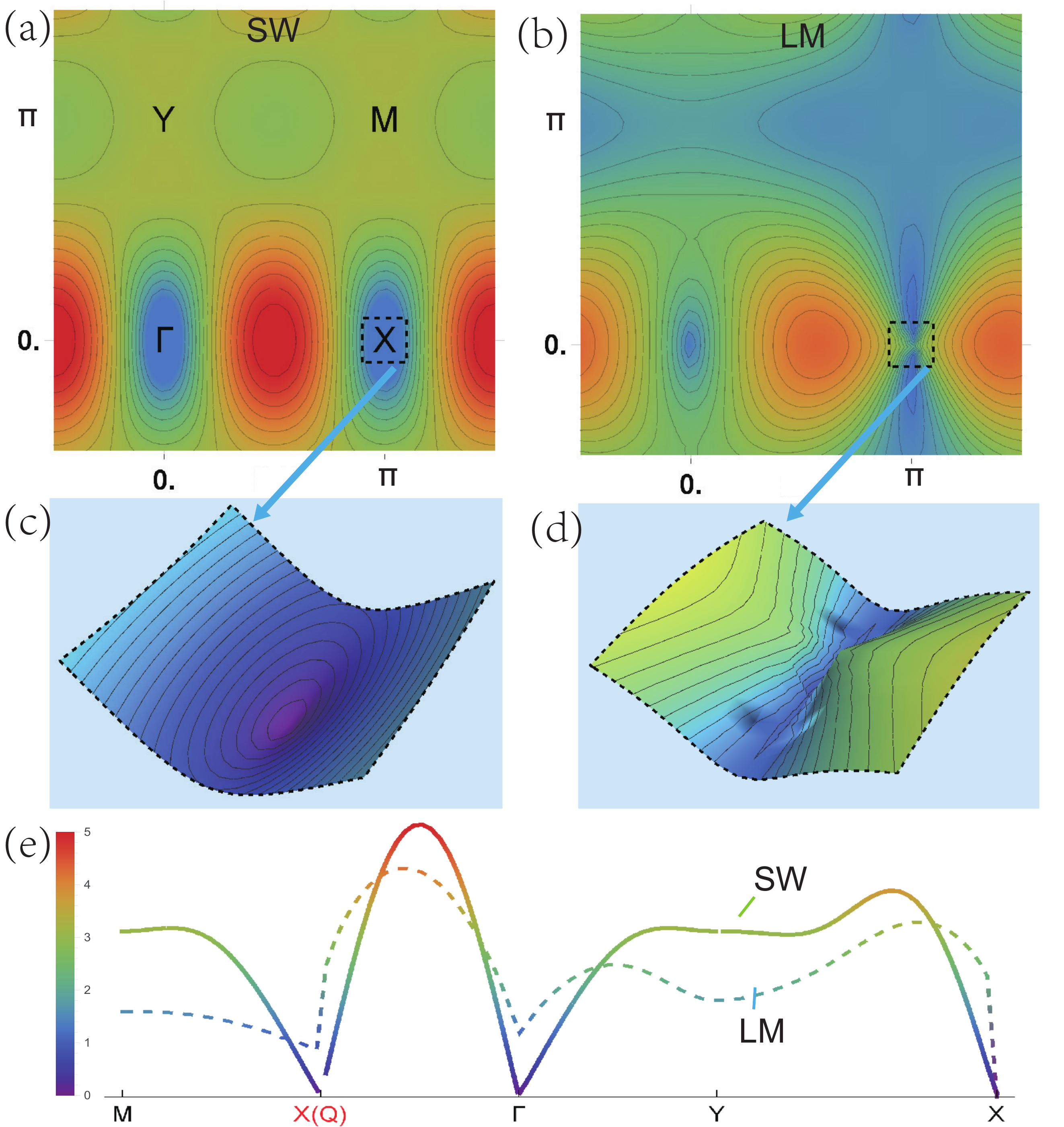}
	\caption{The spectra of SW and LM for CAF phase. Here we used the parameters for $FeSe$ at $9GPa$\cite{Glasbrenner-NP15}: $K=0.2,\ J_2=0.69,\ J_3=0.09$. $A=1.0001$ is used to slightly gap the SW Goldstone modes and speed up the calculation. The color (bar in (e)) marks the energy for SW (a,c)  and LM (b,d).  $\Gamma XMY$ in (a) mark the high symmetry points.   The 3D dispersion (c,d) are taken from the dashed square region $(\pi\pm 0.1\pi,\pm 0.1\pi)$. The SW is gapless ($0.07J_1$) and slightly anisotropic with oval-shape equal-energy slice and that the LM is gapped ($0.87J_1$) and strongly anisotropy with open equal-energy curve bending outward. The projected dispersion (e) clearly shows Goldstone mode and Higgs mode at $X(\mathbf{Q})$.}\label{f:CAF-sp}
 \end{figure}

The spectra of SW and LM for  FeSe at $9\ GPa$\citep{Glasbrenner-NP15} are depicted in  Fig.\ref{f:CAF-sp}, in the unit of $J_1s$ ($51.1$ meV). The Goldstone modes appear at $\Gamma$ and $\mathbf{Q}$ in the SW spectrum Fig.\ref{f:CAF-sp}(a). In our calculation, the anisotropic exchange opens a tiny gap $0.07J_1s$ $(\sim \sqrt{\delta_a})$. This is often introduced to explain the observed anisotropic gap in the transverse modes\citep{Wang-PRX13}. In totally isotropic case $(\delta_a=K=0)$, it would be gapless at all high symmetry points  $YMX\Gamma $ as one can see from Eq.(\ref{e:solve}). Zooming into $\mathbf{Q}$, the SW has oval-shape equal energy line as seen from the 3D view in Fig.\ref{f:CAF-sp}(c) which has been measured experimentally.

The LM spectrum is depicted in  Fig.\ref{f:CAF-sp}(b). It has a deep  valley structure along the $k_x=\pi$ line.  The dispersion is flat in $(\pi,\pm \delta_q)$ direction, but it is steep in $(\pi\pm \delta_q,0)$ as shown in Fig.\ref{f:CAF-sp}(b,d). 
The valley structure is manifested   in Fig.\ref{f:CAF-sp}(d). 
It is interesting to point out that at $\mathbf{Q}=(\pi,0)$,  the structure factor $S(\mathbf{Q})=0$. This is protected by the symmetry $\gamma_{k+Q}=-\gamma_k$, not an issue of approximation.   The coupling in $c$ axis can break this symmetry, resulting in a well-defined LM. Indeed, $J_c$ is important for the occurrence of long range magnetic order. Yet, for numerical consideration, the longitudinal gap can be obtained from the nearby region since Eq.(\ref{EqSq}) is a smooth function except at $\mathbf{Q}$. 

The dispersion   along some high symmetry directions are shown in Fig.\ref{f:CAF-sp}(e). As expected, the transverse mode converges to almost zero and its linear dispersion shows slight anisotropy at $\mathbf{Q}$. The LM at this point is gapped and strongly anisotropic. The dispersion is flat in $q_y$ ($XM$) direction and sharp in $q_x$ ($X\Gamma $) direction. The calculated longitudinal gap  $\Delta =0.87J_1s=44.5$ meV, which is not measured yet.

\begin{figure}[ht] 
\centering 
	\includegraphics[width=.45\textwidth]{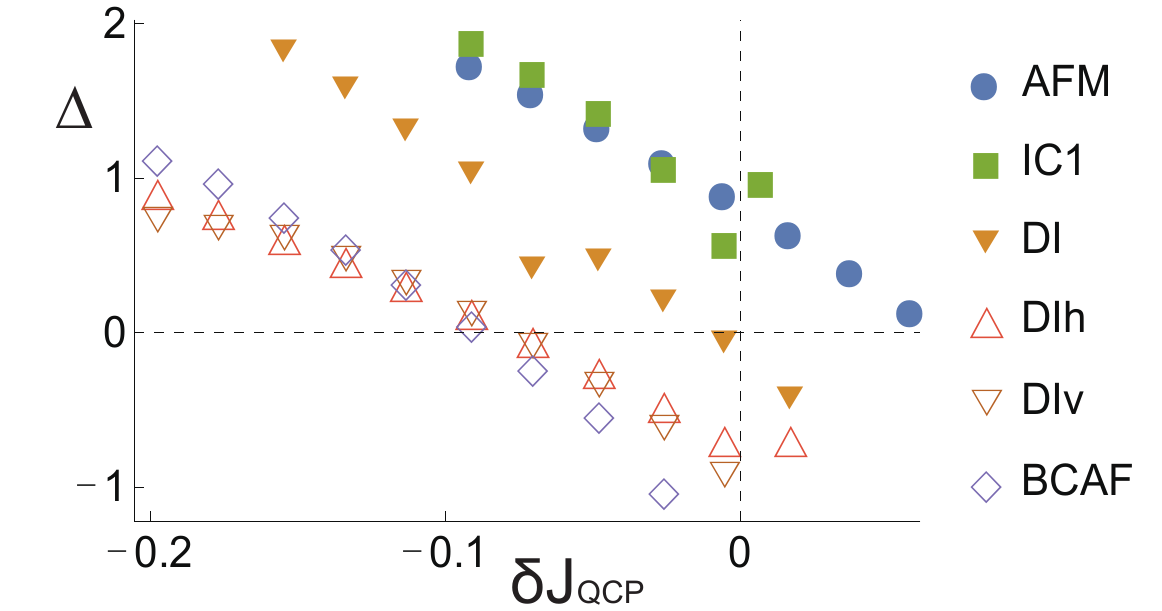}
	\caption{The longitudinal gaps $\Delta$ as functions of the distance $\delta J_{qcp}$ to various neighbor phases. The approaching paths are marked by color arrows  in Fig.\ref{f:phases}(a). The  empty markers $(K=0.5)$ manifest smaller gaps, possibly indicating stronger frustration compared with filled markers $(K=0.2)$.}\label{f:qcp}
 \end{figure}
 
As approaching the phases boundaries, the order parameter would be soften due to the competition among various magnetic orders\citep{affleck-prb92} and  the LM would appear in low energy as frustration arises. The longitudinal gap $\Delta $  is depicted in Fig,\ref{f:qcp} as approaching the phase boundaries.  The gap is taken as the limit  from the flat $k_y$ direction.  The gap $\Delta$ drops to zero near the phases boundaries, appearing as low energy excitation in the neutron scattering experiment. We also noticed that the gaps of the three empty markers (K=0.5), as routes approaching BCAF, DI-horizontally and DI-vertically respectively, drop faster to zero in front of the boundaries, as a possible sign of stronger frustration. No big difference is found for the paths approaching the DI phases horizontally (varying $J_2$) and vertically (varying $J_3$). Thus in this effective model, localized exchange and itinerant coupling equivalently contribute to frustration to hold LM. \\ 

Adjust to the models and parameters in Ref.\citep{Wang-PRX13},  a $53$ meV longitudinal gap can be obtained with our calculation, comparing with the observed value $25\sim 30$ meV. This result is very reasonable. As fluctuation effect in the vicinity of critical region is underestimated in our method, the gap in our calculation should be larger than the true gap. For more accurate quantitative result, the higher order correction has to be considered.   

\subsection{\label{ss:other}Other Magnetic Phase  }
\begin{figure}[ht]
\centering
\includegraphics[width=.45\textwidth]{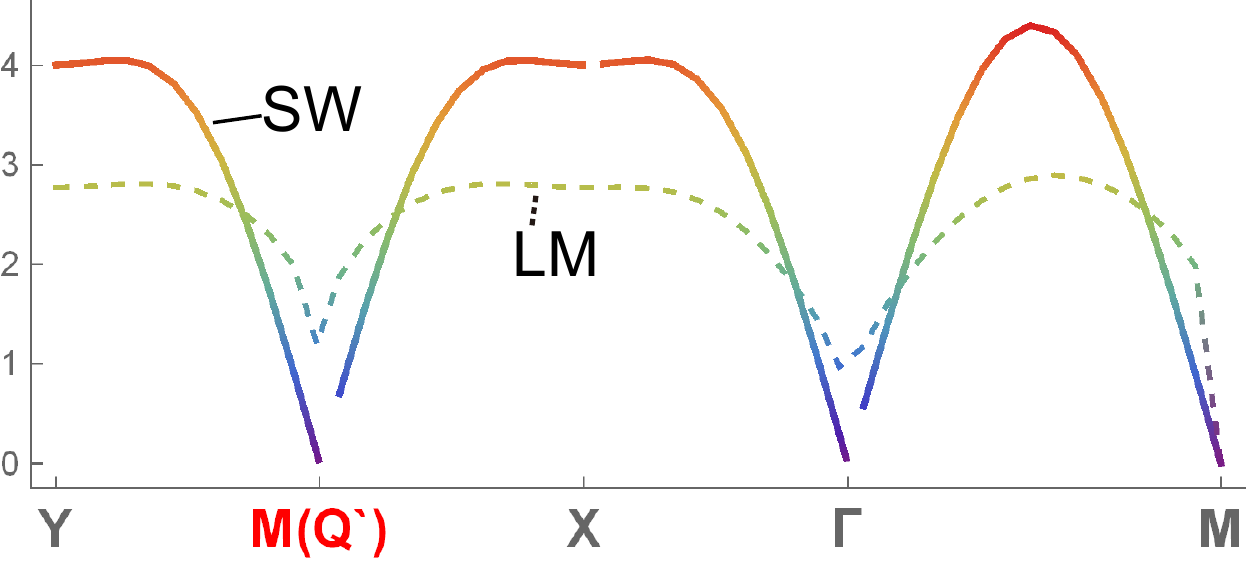}
\caption{The spectra for SW and LM for AF state along the high symmetry points with parameters $K=0.2,\ J_2=0.2,\ J_3=0.05$ and $A=1.0001$. The SW gap at $\mathbf{Q}^\prime$ is zero and LM $\Delta=1.20$ in the unit of $J_1s$.}
\label{f:af-long}
\end{figure}

Similar work can be done for various magnetic phases represented in Fig.\ref{f:phases}, following the method generally described in Appendix.(\ref{a:derivation}). The $J_1$ dominated AF state with $\mathbf{Q}^\prime=M=(\pi,\pi)$ is another example of exactly solvable case. Its SW the LM along the high symmetry points are in Fig.\ref{f:af-long}.  Similar to the CAF phase, the SW is gapless and the LM is gapped at $\mathbf{Q}^\prime$.  The magnetic environment is $C_4$ symmetric and both the SW and LM dispersion are isotropic.  Approaching to the phases boundaries, the longitudinal gaps also decreases to appear as low excitations due to phases competition. The LM are always more sensitive to the quantum frustration. Near the quantum critical point, the interaction among magnons in LM breaks the ground state. 

\section{Discussion and Summary}
 Recently, the inelastic neutron scattering experiments have already measured the LM in iron-based superconducting materials\cite{Wang-PRX13,Luo-Hu-PRL13,wasser-16,zhang-prb14,steffens-prl13}. These  results are typically considered as evidence to support pure itinerant magnetism in these materials\cite{Wang-PRX13,Eremin-prb10,you-prb11}.  However, as argued in literature\citep{Hu-PRB12}, a pure itinerant magnetism can not explain all the observed magnetic properties in iron-based superconductors. According to our calculation, LM can also emerge from the effective exchange model.
 
Combining with the experimental results, our results  strongly support that iron-based superconductors are strongly frustrated magnetic systems in a vicinity to many different magnetic phases. In particular,   the CAF order is very close to  quantum critical transitions to other magnetic orders.  The nature of the frustration stems from the competitions between short range local magnetic exchange couplings and other effective magnetic interactions through couplings with itinerant electrons.  

In summary, we derived the longitudinal excitation for the general $J_1-J_2-J_3-K$ magnetic model. Specifically, the analytic solution in the CAF state is given. The dispersion of the LM in the CAF state near the quantum critical point is very different from the transverse SW modes.  

{\it Acknowledgement:} the work is supported by the Ministry of Science and Technology of China 973 program(Grant No. 2015CB921300), National Science Foundation of China (Grant No. NSFC-11334012, No. NSFC-11534014),  the Strategic Priority Research Program of CAS (Grant No. XDB07000000) and the International Young Scientist Fellowship of Institute of Physics CAS (Grant No.2017002).

\appendix
\section{Derivation}\label{a:derivation}
\subsection{Spin Wave spectrum}
The magnon satisfies the bosonic commute relation:
\begin{eqnarray}
\Psi_k\Psi^\dagger_q-((\Psi^\dagger_q)^T\Psi^T_k)^T=\delta_{kq}\sigma_{zn}.
\end{eqnarray}
With the help of \textit{Bogoliubov} transformation $\Psi_k=T_k\Phi_k$, where $\Phi^\dagger_k=(\mathbf{d}^\dagger_k,\mathbf{d}^T_{-k})$, the Hamiltonian can be diagonalized as $T^\dagger_kH_kT_k=\sigma_{zn}\Lambda_k$, when the transformation matrix satisfy
\begin{eqnarray}
\begin{array}{rl}
\sigma_{zn}=&T_k\sigma_{zn}T^\dagger_k,\label{e:commute}\\
\Lambda_k=&T_k^{-1}(\sigma_{zn}H_k)T_k,\label{e:diag}
\end{array}
\end{eqnarray}
where $\Lambda_k$ is the diagonalized matrix of $\sigma_{zn}H_k$ by similarity transformation and $T_k$ is the Bogoliubov matrix\cite{xiao-09bogoliubov}:
\begin{eqnarray}
T_k=\left(\begin{array}{cc}
\mu_k & \nu_k\\
\nu^\ast_{-k} & \mu^\ast_{-k}
\end{array}\right),
\end{eqnarray}
where $\mu_k,\nu_k$ are the $n\times n$ \textit{coherent phase} matrices. One can collect the eigenvectors of matrix $\sigma_{zn}H_k$ and then normalize them by equation.(\ref{e:commute}). It is proven that for the eigenvalue $\epsilon_k$ with eigenvector $\mathbf{v}_k$, there exists a dual eigenvalue $-\epsilon^\ast_{-k}$ with eigenvector $\sigma_{xn}\mathbf{v}^\ast_{-k}$. Thus by properly order the eigenvalues and adjust the relative phases, this $T_k$ is the Bogoliubov matrix to diagonalize the Hamiltonian via congruence transformation. In algorithm, there could be phase freedom for those eigenvectors, making the matrix non-Bogoliubov, i.e. $T\rightarrow Te^{i\Theta}$ with $\Theta=diag \{\theta_1,\cdots,\theta_{2n} \}$ the phases matrix. It's easy to remove those phases and, moreover, this phase freedom does not affect the SW spectra and the LM.

In linear SW theory, the Hamiltonian matrix elements are the Fourier transformation of the neighborhood interaction in real space, so $\omega_k,\gamma_k,\epsilon_k,\mu_k,\nu_k$ all satisfy $f^\ast(k)=f(-k)$ and the diagonalized Hamiltonian $\sigma_{zn}\Lambda_k$ is real and doubly degenerate.\\

\subsection{Longitudinal Spectrum}
The essence of calculating the ground state expectation of the commutators $N_m(q)$ and $N^K_m(q)$ in equ.(\ref{e:commutator}) is to calculate the correlation of spins, up to second $\langle SS\rangle$ and forth order $\langle SSSS\rangle$. Define the magnon correlators first:
\begin{eqnarray*}
\langle \Psi_k\Psi^\dagger_q\rangle=\left(\begin{array}{cc}
\mu_k\mu^\dagger_k  & \mu_k\nu^T_{-k}\\
\nu^\ast_{-k}\mu^\dagger_k & \nu^\ast_{-k}\nu^T_{-k}
\end{array}\right)\delta_{kq}=\left(\begin{array}{cc}
\rho^T_k+1 & \Delta_{-k}\\
\Delta^\dagger_{-k} & \rho_{-k}
\end{array}\right)\delta_{kq}.
\end{eqnarray*}
It is easy to see $\rho^\dagger_k=\rho_k$ and $\Delta_k=\Delta^T_{-k}$. In linear spin theory, $f(-k)=f^\ast(k)$ also works for $\rho_k$ and $\Delta_k$. The real space correlations 
\begin{eqnarray}\label{tpc}
\rho(m)&\equiv \langle \mathbf{a}^{\dagger,T}_l\mathbf{a}^T_{l+m}\rangle=\frac 1{N^\prime}\sum_k e^{-ikm}\rho_k,\\
\Delta(m)&\equiv \langle \mathbf{a}_l\mathbf{a}^T_{l+m}\rangle=\frac 1{N^\prime}\sum_k e^{-ikm}\Delta_k
\end{eqnarray}
satisfy $\rho^\ast(m)=\rho^\dagger(-m)=\rho(m)$ and $\Delta^\ast(m)=\Delta^\dagger(-m)=\Delta(m)$. Thus the correlation among different magnons is the elements of the correlation matrices $\rho(m)$ and $\Delta(m)$:
\begin{eqnarray}
\begin{array}{rl}
\langle a^\dagger_lb_{l+m}\rangle=&\rho_{ab}(m)= \langle a_lb^\dagger_{l+m}\rangle-\delta_{m0},\\
\langle a_lb_{l+m}\rangle=&\Delta_{ab}(m)=\langle a^\dagger_lb^\dagger_{l+m}\rangle.
\end{array}
\end{eqnarray}
For higher (even) operators correlation, thereafter, the \textit{contraction rules} can be concluded $\langle c_1c_2\cdots c_{2n-1} c_{2n}\rangle$ ($c_i=a_i,a_i^\dagger$): 
\begin{enumerate}
\item Put the $c's$ operators in pairs, all possible combinations;
\item Refer to the matrix element of $\rho(m)$ and $\Delta(m)$, write all the operator pairs as two-operator correlation functions in real space.
\end{enumerate}
Referring this contraction rule, the following four operators (and their conjugate) correlation is needed:
\begin{eqnarray}
\begin{array}{rl}
\langle a_ib_ja_ib_j\rangle=&2\Delta_{ab}^2+\Delta_{aa}\Delta_{bb},\\
\langle a_ib^\dagger_ja_ib^\dagger_j\rangle=&2\rho_{ab}^2+\Delta_{aa}\Delta_{bb},\\
\langle a^\dagger_ib^\dagger_ja^\dagger_ia_j\rangle=&2\rho_{aa}\Delta_{ab}+\rho_{ab}\Delta_{aa},\\
\langle a_ib^\dagger_jb^\dagger_jb_j\rangle=&=2\rho_{bb}\rho_{ab}+\Delta_{ab}\Delta_{bb}.
\end{array}
\end{eqnarray}
We omit the variables $(j-i)$ within $\rho_{ab},\Delta_{ab}$ for convenience. Relevant to the LM up to the four operators correlation, for $N_m(q)$, we need to estimate :
\begin{eqnarray*}
\langle S^\dagger_i S^-_j\rangle=& 2s\rho_{ab}-\rho_{ab}(\rho_{aa}+\rho_{bb})-\Delta_{ab}(\Delta_{aa}+\Delta_{bb})/2,\\
\langle S^-_i S^-_j\rangle=& 2s\Delta_{ab}-\Delta_{ab}(\rho_{aa}+\rho_{bb})-\rho_{ab}(\Delta_{aa}+\Delta_{bb})/2,
\end{eqnarray*}
and in the quartic term $N^K_m(q)$
\begin{eqnarray}
\begin{array}{rl}
\langle S^-_iS^-_jS^-_iS^-_j\rangle \approx & (2s)^2(2\Delta_{ab}^2+\Delta_{aa}\Delta_{bb}),\\
\langle S^\dagger_i S^-_jS^\dagger_i S^-_j\rangle\approx &(2s)^2(2\rho_{ab}^2+\Delta_{aa}\Delta_{bb}),\\
\langle S^\dagger_i S^-_jS^z_iS^z_j\rangle \approx & s^2\bigl(2(s-1)\rho_{ab}-5\rho_{ab}(\rho_{aa}+\rho_{bb})\nonumber\\\ 
&-5/2\Delta_{ab}(\Delta_{aa}+\Delta_{bb})\bigr),\\
\langle S^\dagger_i S^\dagger_j S^z_iS^z_j\rangle\approx & s^2\bigl(2(s-2)\Delta_{ab}-5\Delta_{ab}(\rho_{aa}+\rho_{bb})\nonumber\\\ 
&-5/2\rho_{ab}(\Delta_{aa}+\Delta_{bb})\bigr),\\
\end{array}
\end{eqnarray}
With the above correlators derived, the correlation function $\Pi$'s in $N(q)$ can be obtained.
\\

\bibliography{longitudinal-bib}

\begin{thebibliography}{35}
\expandafter\ifx\csname natexlab\endcsname\relax\def\natexlab#1{#1}\fi
\expandafter\ifx\csname bibnamefont\endcsname\relax
  \def\bibnamefont#1{#1}\fi
\expandafter\ifx\csname bibfnamefont\endcsname\relax
  \def\bibfnamefont#1{#1}\fi
\expandafter\ifx\csname citenamefont\endcsname\relax
  \def\citenamefont#1{#1}\fi
\expandafter\ifx\csname url\endcsname\relax
  \def\url#1{\texttt{#1}}\fi
\expandafter\ifx\csname urlprefix\endcsname\relax\def\urlprefix{URL }\fi
\providecommand{\bibinfo}[2]{#2}
\providecommand{\eprint}[2][]{\url{#2}}

\bibitem[{\citenamefont{Dai et~al.}(2012)\citenamefont{Dai, Hu, and
  Dagotto}}]{dai-np12}
\bibinfo{author}{\bibfnamefont{P.}~\bibnamefont{Dai}},
  \bibinfo{author}{\bibfnamefont{J.}~\bibnamefont{Hu}}, \bibnamefont{and}
  \bibinfo{author}{\bibfnamefont{E.}~\bibnamefont{Dagotto}},
  \bibinfo{journal}{Nature Physics} \textbf{\bibinfo{volume}{8}},
  \bibinfo{pages}{709} (\bibinfo{year}{2012}).

\bibitem[{\citenamefont{Dai}(2015)}]{Dai-RMP15}
\bibinfo{author}{\bibfnamefont{P.}~\bibnamefont{Dai}}, \bibinfo{journal}{Rev.
  Mod. Phys.} \textbf{\bibinfo{volume}{87}}, \bibinfo{pages}{855}
  (\bibinfo{year}{2015}),
  \urlprefix\url{http://link.aps.org/doi/10.1103/RevModPhys.87.855}.

\bibitem[{\citenamefont{de~La~Cruz et~al.}(2008)\citenamefont{de~La~Cruz,
  Huang, Lynn, Li, Ratcliff~Ii, Zarestky, Mook, Chen, Luo, Wang
  et~al.}}]{de-na08}
\bibinfo{author}{\bibfnamefont{C.}~\bibnamefont{de~La~Cruz}},
  \bibinfo{author}{\bibfnamefont{Q.}~\bibnamefont{Huang}},
  \bibinfo{author}{\bibfnamefont{J.}~\bibnamefont{Lynn}},
  \bibinfo{author}{\bibfnamefont{J.}~\bibnamefont{Li}},
  \bibinfo{author}{\bibfnamefont{W.}~\bibnamefont{Ratcliff~Ii}},
  \bibinfo{author}{\bibfnamefont{J.~L.} \bibnamefont{Zarestky}},
  \bibinfo{author}{\bibfnamefont{H.}~\bibnamefont{Mook}},
  \bibinfo{author}{\bibfnamefont{G.}~\bibnamefont{Chen}},
  \bibinfo{author}{\bibfnamefont{J.}~\bibnamefont{Luo}},
  \bibinfo{author}{\bibfnamefont{N.}~\bibnamefont{Wang}}, \bibnamefont{et~al.},
  \bibinfo{journal}{nature} \textbf{\bibinfo{volume}{453}},
  \bibinfo{pages}{899} (\bibinfo{year}{2008}).

\bibitem[{\citenamefont{Marty et~al.}(2011)\citenamefont{Marty, Christianson,
  Wang, Matsuda, Cao, VanBebber, Zarestky, Singh, Sefat, and
  Lumsden}}]{Marty-PRB11}
\bibinfo{author}{\bibfnamefont{K.}~\bibnamefont{Marty}},
  \bibinfo{author}{\bibfnamefont{A.~D.} \bibnamefont{Christianson}},
  \bibinfo{author}{\bibfnamefont{C.~H.} \bibnamefont{Wang}},
  \bibinfo{author}{\bibfnamefont{M.}~\bibnamefont{Matsuda}},
  \bibinfo{author}{\bibfnamefont{H.}~\bibnamefont{Cao}},
  \bibinfo{author}{\bibfnamefont{L.~H.} \bibnamefont{VanBebber}},
  \bibinfo{author}{\bibfnamefont{J.~L.} \bibnamefont{Zarestky}},
  \bibinfo{author}{\bibfnamefont{D.~J.} \bibnamefont{Singh}},
  \bibinfo{author}{\bibfnamefont{A.~S.} \bibnamefont{Sefat}}, \bibnamefont{and}
  \bibinfo{author}{\bibfnamefont{M.~D.} \bibnamefont{Lumsden}},
  \bibinfo{journal}{Phys. Rev. B} \textbf{\bibinfo{volume}{83}},
  \bibinfo{pages}{060509} (\bibinfo{year}{2011}),
  \urlprefix\url{http://link.aps.org/doi/10.1103/PhysRevB.83.060509}.

\bibitem[{\citenamefont{Bao et~al.}(2009)\citenamefont{Bao, Qiu, Huang, Green,
  Zajdel, Fitzsimmons, Zhernenkov, Chang, Fang, Qian et~al.}}]{bao-prl09}
\bibinfo{author}{\bibfnamefont{W.}~\bibnamefont{Bao}},
  \bibinfo{author}{\bibfnamefont{Y.}~\bibnamefont{Qiu}},
  \bibinfo{author}{\bibfnamefont{Q.}~\bibnamefont{Huang}},
  \bibinfo{author}{\bibfnamefont{M.~A.} \bibnamefont{Green}},
  \bibinfo{author}{\bibfnamefont{P.}~\bibnamefont{Zajdel}},
  \bibinfo{author}{\bibfnamefont{M.~R.} \bibnamefont{Fitzsimmons}},
  \bibinfo{author}{\bibfnamefont{M.}~\bibnamefont{Zhernenkov}},
  \bibinfo{author}{\bibfnamefont{S.}~\bibnamefont{Chang}},
  \bibinfo{author}{\bibfnamefont{M.}~\bibnamefont{Fang}},
  \bibinfo{author}{\bibfnamefont{B.}~\bibnamefont{Qian}}, \bibnamefont{et~al.},
  \bibinfo{journal}{Phys. Rev. Lett.} \textbf{\bibinfo{volume}{102}},
  \bibinfo{pages}{247001} (\bibinfo{year}{2009}),
  \urlprefix\url{http://link.aps.org/doi/10.1103/PhysRevLett.102.247001}.

\bibitem[{\citenamefont{Cao et~al.}(2015)\citenamefont{Cao, Chen, Xiang, and
  Gong}}]{cao-prb15}
\bibinfo{author}{\bibfnamefont{H.-Y.} \bibnamefont{Cao}},
  \bibinfo{author}{\bibfnamefont{S.}~\bibnamefont{Chen}},
  \bibinfo{author}{\bibfnamefont{H.}~\bibnamefont{Xiang}}, \bibnamefont{and}
  \bibinfo{author}{\bibfnamefont{X.-G.} \bibnamefont{Gong}},
  \bibinfo{journal}{Physical Review B} \textbf{\bibinfo{volume}{91}},
  \bibinfo{pages}{020504} (\bibinfo{year}{2015}).

\bibitem[{\citenamefont{Mazin et~al.}(2008)\citenamefont{Mazin, Singh,
  Johannes, and Du}}]{mazin-prl08}
\bibinfo{author}{\bibfnamefont{I.~I.} \bibnamefont{Mazin}},
  \bibinfo{author}{\bibfnamefont{D.~J.} \bibnamefont{Singh}},
  \bibinfo{author}{\bibfnamefont{M.~D.} \bibnamefont{Johannes}},
  \bibnamefont{and} \bibinfo{author}{\bibfnamefont{M.~H.} \bibnamefont{Du}},
  \bibinfo{journal}{Phys. Rev. Lett.} \textbf{\bibinfo{volume}{101}},
  \bibinfo{pages}{057003} (\bibinfo{year}{2008}),
  \urlprefix\url{http://link.aps.org/doi/10.1103/PhysRevLett.101.057003}.

\bibitem[{\citenamefont{Hu et~al.}(2012)\citenamefont{Hu, Xu, Liu, Hao, and
  Wang}}]{Hu-PRB12}
\bibinfo{author}{\bibfnamefont{J.}~\bibnamefont{Hu}},
  \bibinfo{author}{\bibfnamefont{B.}~\bibnamefont{Xu}},
  \bibinfo{author}{\bibfnamefont{W.}~\bibnamefont{Liu}},
  \bibinfo{author}{\bibfnamefont{N.-N.} \bibnamefont{Hao}}, \bibnamefont{and}
  \bibinfo{author}{\bibfnamefont{Y.}~\bibnamefont{Wang}},
  \bibinfo{journal}{Phys. Rev. B} \textbf{\bibinfo{volume}{85}},
  \bibinfo{pages}{144403} (\bibinfo{year}{2012}),
  \urlprefix\url{http://link.aps.org/doi/10.1103/PhysRevB.85.144403}.

\bibitem[{\citenamefont{Glasbrenner et~al.}(2015)\citenamefont{Glasbrenner,
  Mazin, Jeschke, Hirschfeld, Fernandes, and Valent{\'\i}}}]{Glasbrenner-NP15}
\bibinfo{author}{\bibfnamefont{J.}~\bibnamefont{Glasbrenner}},
  \bibinfo{author}{\bibfnamefont{I.}~\bibnamefont{Mazin}},
  \bibinfo{author}{\bibfnamefont{H.~O.} \bibnamefont{Jeschke}},
  \bibinfo{author}{\bibfnamefont{P.}~\bibnamefont{Hirschfeld}},
  \bibinfo{author}{\bibfnamefont{R.}~\bibnamefont{Fernandes}},
  \bibnamefont{and}
  \bibinfo{author}{\bibfnamefont{R.}~\bibnamefont{Valent{\'\i}}},
  \bibinfo{journal}{Nature Physics} \textbf{\bibinfo{volume}{11}},
  \bibinfo{pages}{953} (\bibinfo{year}{2015}).

\bibitem[{\citenamefont{Fang et~al.}(2009)\citenamefont{Fang, Bernevig, and
  Hu}}]{fang-EPL09}
\bibinfo{author}{\bibfnamefont{C.}~\bibnamefont{Fang}},
  \bibinfo{author}{\bibfnamefont{B.~A.} \bibnamefont{Bernevig}},
  \bibnamefont{and} \bibinfo{author}{\bibfnamefont{J.}~\bibnamefont{Hu}},
  \bibinfo{journal}{EPL (Europhysics Letters)} \textbf{\bibinfo{volume}{86}},
  \bibinfo{pages}{67005} (\bibinfo{year}{2009}).

\bibitem[{\citenamefont{Fang et~al.}(2008)\citenamefont{Fang, Yao, Tsai, Hu,
  and Kivelson}}]{fang-prb08}
\bibinfo{author}{\bibfnamefont{C.}~\bibnamefont{Fang}},
  \bibinfo{author}{\bibfnamefont{H.}~\bibnamefont{Yao}},
  \bibinfo{author}{\bibfnamefont{W.-F.} \bibnamefont{Tsai}},
  \bibinfo{author}{\bibfnamefont{J.}~\bibnamefont{Hu}}, \bibnamefont{and}
  \bibinfo{author}{\bibfnamefont{S.~A.} \bibnamefont{Kivelson}},
  \bibinfo{journal}{Physical Review B} \textbf{\bibinfo{volume}{77}},
  \bibinfo{pages}{224509} (\bibinfo{year}{2008}).

\bibitem[{\citenamefont{Yaresko et~al.}(2009)\citenamefont{Yaresko, Liu,
  Antonov, and Andersen}}]{prb09-kpaper}
\bibinfo{author}{\bibfnamefont{A.~N.} \bibnamefont{Yaresko}},
  \bibinfo{author}{\bibfnamefont{G.-Q.} \bibnamefont{Liu}},
  \bibinfo{author}{\bibfnamefont{V.~N.} \bibnamefont{Antonov}},
  \bibnamefont{and} \bibinfo{author}{\bibfnamefont{O.~K.}
  \bibnamefont{Andersen}}, \bibinfo{journal}{Phys. Rev. B}
  \textbf{\bibinfo{volume}{79}}, \bibinfo{pages}{144421}
  (\bibinfo{year}{2009}),
  \urlprefix\url{http://link.aps.org/doi/10.1103/PhysRevB.79.144421}.

\bibitem[{\citenamefont{Wysocki et~al.}(2011)\citenamefont{Wysocki,
  Belashchenko, and Antropov}}]{wysocki-np11}
\bibinfo{author}{\bibfnamefont{A.~L.} \bibnamefont{Wysocki}},
  \bibinfo{author}{\bibfnamefont{K.~D.} \bibnamefont{Belashchenko}},
  \bibnamefont{and} \bibinfo{author}{\bibfnamefont{V.~P.}
  \bibnamefont{Antropov}}, \bibinfo{journal}{Nature Physics}
  \textbf{\bibinfo{volume}{7}}, \bibinfo{pages}{485} (\bibinfo{year}{2011}).

\bibitem[{\citenamefont{Glasbrenner et~al.}(2014)\citenamefont{Glasbrenner,
  Velev, and Mazin}}]{mazin-prb08}
\bibinfo{author}{\bibfnamefont{J.~K.} \bibnamefont{Glasbrenner}},
  \bibinfo{author}{\bibfnamefont{J.~P.} \bibnamefont{Velev}}, \bibnamefont{and}
  \bibinfo{author}{\bibfnamefont{I.~I.} \bibnamefont{Mazin}},
  \bibinfo{journal}{Phys. Rev. B} \textbf{\bibinfo{volume}{89}},
  \bibinfo{pages}{064509} (\bibinfo{year}{2014}),
  \urlprefix\url{http://link.aps.org/doi/10.1103/PhysRevB.89.064509}.

\bibitem[{\citenamefont{Wang et~al.}(2011)\citenamefont{Wang, Fang, Yao, Tan,
  Harriger, Song, Netherton, Zhang, Wang, Stone et~al.}}]{wang-nc11}
\bibinfo{author}{\bibfnamefont{M.}~\bibnamefont{Wang}},
  \bibinfo{author}{\bibfnamefont{C.}~\bibnamefont{Fang}},
  \bibinfo{author}{\bibfnamefont{D.-X.} \bibnamefont{Yao}},
  \bibinfo{author}{\bibfnamefont{G.}~\bibnamefont{Tan}},
  \bibinfo{author}{\bibfnamefont{L.~W.} \bibnamefont{Harriger}},
  \bibinfo{author}{\bibfnamefont{Y.}~\bibnamefont{Song}},
  \bibinfo{author}{\bibfnamefont{T.}~\bibnamefont{Netherton}},
  \bibinfo{author}{\bibfnamefont{C.}~\bibnamefont{Zhang}},
  \bibinfo{author}{\bibfnamefont{M.}~\bibnamefont{Wang}},
  \bibinfo{author}{\bibfnamefont{M.~B.} \bibnamefont{Stone}},
  \bibnamefont{et~al.}, \bibinfo{journal}{Nature communications}
  \textbf{\bibinfo{volume}{2}}, \bibinfo{pages}{580} (\bibinfo{year}{2011}).

\bibitem[{\citenamefont{Giovannetti et~al.}(2011)\citenamefont{Giovannetti,
  Ortix, Marsman, Capone, Van Den~Brink, and Lorenzana}}]{giovannetti-nc11}
\bibinfo{author}{\bibfnamefont{G.}~\bibnamefont{Giovannetti}},
  \bibinfo{author}{\bibfnamefont{C.}~\bibnamefont{Ortix}},
  \bibinfo{author}{\bibfnamefont{M.}~\bibnamefont{Marsman}},
  \bibinfo{author}{\bibfnamefont{M.}~\bibnamefont{Capone}},
  \bibinfo{author}{\bibfnamefont{J.}~\bibnamefont{Van Den~Brink}},
  \bibnamefont{and}
  \bibinfo{author}{\bibfnamefont{J.}~\bibnamefont{Lorenzana}},
  \bibinfo{journal}{Nature communications} \textbf{\bibinfo{volume}{2}},
  \bibinfo{pages}{398} (\bibinfo{year}{2011}).

\bibitem[{\citenamefont{Hoyer et~al.}(2016)\citenamefont{Hoyer, Fernandes,
  Levchenko, and Schmalian}}]{jorg-prb16}
\bibinfo{author}{\bibfnamefont{M.}~\bibnamefont{Hoyer}},
  \bibinfo{author}{\bibfnamefont{R.~M.} \bibnamefont{Fernandes}},
  \bibinfo{author}{\bibfnamefont{A.}~\bibnamefont{Levchenko}},
  \bibnamefont{and}
  \bibinfo{author}{\bibfnamefont{J.}~\bibnamefont{Schmalian}},
  \bibinfo{journal}{Phys. Rev. B} \textbf{\bibinfo{volume}{93}},
  \bibinfo{pages}{144414} (\bibinfo{year}{2016}),
  \urlprefix\url{https://link.aps.org/doi/10.1103/PhysRevB.93.144414}.

\bibitem[{\citenamefont{Haldane}(1983)}]{haldane-prl83}
\bibinfo{author}{\bibfnamefont{F.~D.~M.} \bibnamefont{Haldane}},
  \bibinfo{journal}{Phys. Rev. Lett.} \textbf{\bibinfo{volume}{50}},
  \bibinfo{pages}{1153} (\bibinfo{year}{1983}),
  \urlprefix\url{http://link.aps.org/doi/10.1103/PhysRevLett.50.1153}.

\bibitem[{\citenamefont{Wang et~al.}(2013)\citenamefont{Wang, Zhang, Wang, Luo,
  Regnault, Dai, and Li}}]{Wang-PRX13}
\bibinfo{author}{\bibfnamefont{C.}~\bibnamefont{Wang}},
  \bibinfo{author}{\bibfnamefont{R.}~\bibnamefont{Zhang}},
  \bibinfo{author}{\bibfnamefont{F.}~\bibnamefont{Wang}},
  \bibinfo{author}{\bibfnamefont{H.}~\bibnamefont{Luo}},
  \bibinfo{author}{\bibfnamefont{L.~P.} \bibnamefont{Regnault}},
  \bibinfo{author}{\bibfnamefont{P.}~\bibnamefont{Dai}}, \bibnamefont{and}
  \bibinfo{author}{\bibfnamefont{Y.}~\bibnamefont{Li}}, \bibinfo{journal}{Phys.
  Rev. X} \textbf{\bibinfo{volume}{3}}, \bibinfo{pages}{041036}
  (\bibinfo{year}{2013}),
  \urlprefix\url{http://link.aps.org/doi/10.1103/PhysRevX.3.041036}.

\bibitem[{\citenamefont{Luo et~al.}(2013)\citenamefont{Luo, Wang, Zhang, Lu,
  Regnault, Zhang, Li, Hu, and Dai}}]{Luo-Hu-PRL13}
\bibinfo{author}{\bibfnamefont{H.}~\bibnamefont{Luo}},
  \bibinfo{author}{\bibfnamefont{M.}~\bibnamefont{Wang}},
  \bibinfo{author}{\bibfnamefont{C.}~\bibnamefont{Zhang}},
  \bibinfo{author}{\bibfnamefont{X.}~\bibnamefont{Lu}},
  \bibinfo{author}{\bibfnamefont{L.-P.} \bibnamefont{Regnault}},
  \bibinfo{author}{\bibfnamefont{R.}~\bibnamefont{Zhang}},
  \bibinfo{author}{\bibfnamefont{S.}~\bibnamefont{Li}},
  \bibinfo{author}{\bibfnamefont{J.}~\bibnamefont{Hu}}, \bibnamefont{and}
  \bibinfo{author}{\bibfnamefont{P.}~\bibnamefont{Dai}},
  \bibinfo{journal}{Phys. Rev. Lett.} \textbf{\bibinfo{volume}{111}},
  \bibinfo{pages}{107006} (\bibinfo{year}{2013}),
  \urlprefix\url{http://link.aps.org/doi/10.1103/PhysRevLett.111.107006}.

\bibitem[{\citenamefont{Qureshi et~al.}(2012)\citenamefont{Qureshi, Steffens,
  Wurmehl, Aswartham, B\"uchner, and Braden}}]{qureshi-prb12}
\bibinfo{author}{\bibfnamefont{N.}~\bibnamefont{Qureshi}},
  \bibinfo{author}{\bibfnamefont{P.}~\bibnamefont{Steffens}},
  \bibinfo{author}{\bibfnamefont{S.}~\bibnamefont{Wurmehl}},
  \bibinfo{author}{\bibfnamefont{S.}~\bibnamefont{Aswartham}},
  \bibinfo{author}{\bibfnamefont{B.}~\bibnamefont{B\"uchner}},
  \bibnamefont{and} \bibinfo{author}{\bibfnamefont{M.}~\bibnamefont{Braden}},
  \bibinfo{journal}{Phys. Rev. B} \textbf{\bibinfo{volume}{86}},
  \bibinfo{pages}{060410} (\bibinfo{year}{2012}),
  \urlprefix\url{http://link.aps.org/doi/10.1103/PhysRevB.86.060410}.

\bibitem[{\citenamefont{Song et~al.}(2013)\citenamefont{Song, Regnault, Zhang,
  Tan, Carr, Chi, Christianson, Xiang, and Dai}}]{song-prb13}
\bibinfo{author}{\bibfnamefont{Y.}~\bibnamefont{Song}},
  \bibinfo{author}{\bibfnamefont{L.-P.} \bibnamefont{Regnault}},
  \bibinfo{author}{\bibfnamefont{C.}~\bibnamefont{Zhang}},
  \bibinfo{author}{\bibfnamefont{G.}~\bibnamefont{Tan}},
  \bibinfo{author}{\bibfnamefont{S.~V.} \bibnamefont{Carr}},
  \bibinfo{author}{\bibfnamefont{S.}~\bibnamefont{Chi}},
  \bibinfo{author}{\bibfnamefont{A.~D.} \bibnamefont{Christianson}},
  \bibinfo{author}{\bibfnamefont{T.}~\bibnamefont{Xiang}}, \bibnamefont{and}
  \bibinfo{author}{\bibfnamefont{P.}~\bibnamefont{Dai}},
  \bibinfo{journal}{Phys. Rev. B} \textbf{\bibinfo{volume}{88}},
  \bibinfo{pages}{134512} (\bibinfo{year}{2013}),
  \urlprefix\url{http://link.aps.org/doi/10.1103/PhysRevB.88.134512}.

\bibitem[{\citenamefont{Li et~al.}(2017)\citenamefont{Li, Wang, Song, Man, Lu,
  Bourdarot, and Dai}}]{dai-prb17}
\bibinfo{author}{\bibfnamefont{Y.}~\bibnamefont{Li}},
  \bibinfo{author}{\bibfnamefont{W.}~\bibnamefont{Wang}},
  \bibinfo{author}{\bibfnamefont{Y.}~\bibnamefont{Song}},
  \bibinfo{author}{\bibfnamefont{H.}~\bibnamefont{Man}},
  \bibinfo{author}{\bibfnamefont{X.}~\bibnamefont{Lu}},
  \bibinfo{author}{\bibfnamefont{F.}~\bibnamefont{Bourdarot}},
  \bibnamefont{and} \bibinfo{author}{\bibfnamefont{P.}~\bibnamefont{Dai}},
  \bibinfo{journal}{Phys. Rev. B} \textbf{\bibinfo{volume}{96}},
  \bibinfo{pages}{020404} (\bibinfo{year}{2017}),
  \urlprefix\url{https://link.aps.org/doi/10.1103/PhysRevB.96.020404}.

\bibitem[{\citenamefont{Zaliznyak et~al.}(2015)\citenamefont{Zaliznyak, Savici,
  Lumsden, Tsvelik, Hu, and Petrovic}}]{zaliznyak-15spin}
\bibinfo{author}{\bibfnamefont{I.}~\bibnamefont{Zaliznyak}},
  \bibinfo{author}{\bibfnamefont{A.~T.} \bibnamefont{Savici}},
  \bibinfo{author}{\bibfnamefont{M.}~\bibnamefont{Lumsden}},
  \bibinfo{author}{\bibfnamefont{A.}~\bibnamefont{Tsvelik}},
  \bibinfo{author}{\bibfnamefont{R.}~\bibnamefont{Hu}}, \bibnamefont{and}
  \bibinfo{author}{\bibfnamefont{C.}~\bibnamefont{Petrovic}},
  \bibinfo{journal}{Proceedings of the National Academy of Sciences}
  \textbf{\bibinfo{volume}{112}}, \bibinfo{pages}{10316}
  (\bibinfo{year}{2015}).

\bibitem[{\citenamefont{Zhou et~al.}(2013)\citenamefont{Zhou, Huang, Monney,
  Dai, Strocov, Wang, Chen, Zhang, Dai, Patthey et~al.}}]{zhou-nc13}
\bibinfo{author}{\bibfnamefont{K.-J.} \bibnamefont{Zhou}},
  \bibinfo{author}{\bibfnamefont{Y.-B.} \bibnamefont{Huang}},
  \bibinfo{author}{\bibfnamefont{C.}~\bibnamefont{Monney}},
  \bibinfo{author}{\bibfnamefont{X.}~\bibnamefont{Dai}},
  \bibinfo{author}{\bibfnamefont{V.~N.} \bibnamefont{Strocov}},
  \bibinfo{author}{\bibfnamefont{N.-L.} \bibnamefont{Wang}},
  \bibinfo{author}{\bibfnamefont{Z.-G.} \bibnamefont{Chen}},
  \bibinfo{author}{\bibfnamefont{C.}~\bibnamefont{Zhang}},
  \bibinfo{author}{\bibfnamefont{P.}~\bibnamefont{Dai}},
  \bibinfo{author}{\bibfnamefont{L.}~\bibnamefont{Patthey}},
  \bibnamefont{et~al.}, \bibinfo{journal}{Nature communications}
  \textbf{\bibinfo{volume}{4}}, \bibinfo{pages}{1470} (\bibinfo{year}{2013}).

\bibitem[{\citenamefont{You et~al.}(2011)\citenamefont{You, Yang, Kou, and
  Weng}}]{you-prb11}
\bibinfo{author}{\bibfnamefont{Y.-Z.} \bibnamefont{You}},
  \bibinfo{author}{\bibfnamefont{F.}~\bibnamefont{Yang}},
  \bibinfo{author}{\bibfnamefont{S.-P.} \bibnamefont{Kou}}, \bibnamefont{and}
  \bibinfo{author}{\bibfnamefont{Z.-Y.} \bibnamefont{Weng}},
  \bibinfo{journal}{Physical Review B} \textbf{\bibinfo{volume}{84}},
  \bibinfo{pages}{054527} (\bibinfo{year}{2011}).

\bibitem[{\citenamefont{Knolle et~al.}(2010)\citenamefont{Knolle, Eremin,
  Chubukov, and Moessner}}]{Eremin-prb10}
\bibinfo{author}{\bibfnamefont{J.}~\bibnamefont{Knolle}},
  \bibinfo{author}{\bibfnamefont{I.}~\bibnamefont{Eremin}},
  \bibinfo{author}{\bibfnamefont{A.}~\bibnamefont{Chubukov}}, \bibnamefont{and}
  \bibinfo{author}{\bibfnamefont{R.}~\bibnamefont{Moessner}},
  \bibinfo{journal}{Physical Review B} \textbf{\bibinfo{volume}{81}},
  \bibinfo{pages}{140506} (\bibinfo{year}{2010}).

\bibitem[{\citenamefont{Chandra et~al.}(1990)\citenamefont{Chandra, Coleman,
  and Larkin}}]{chandra-prl90}
\bibinfo{author}{\bibfnamefont{P.}~\bibnamefont{Chandra}},
  \bibinfo{author}{\bibfnamefont{P.}~\bibnamefont{Coleman}}, \bibnamefont{and}
  \bibinfo{author}{\bibfnamefont{A.~I.} \bibnamefont{Larkin}},
  \bibinfo{journal}{Phys. Rev. Lett.} \textbf{\bibinfo{volume}{64}},
  \bibinfo{pages}{88} (\bibinfo{year}{1990}),
  \urlprefix\url{http://link.aps.org/doi/10.1103/PhysRevLett.64.88}.

\bibitem[{\citenamefont{Xiao}(2009)}]{xiao-09bogoliubov}
\bibinfo{author}{\bibfnamefont{M.-w.} \bibnamefont{Xiao}},
  \bibinfo{journal}{arXiv preprint arXiv:0908.0787}  (\bibinfo{year}{2009}).

\bibitem[{\citenamefont{Affleck and Wellman}(1992)}]{affleck-prb92}
\bibinfo{author}{\bibfnamefont{I.}~\bibnamefont{Affleck}} \bibnamefont{and}
  \bibinfo{author}{\bibfnamefont{G.~F.} \bibnamefont{Wellman}},
  \bibinfo{journal}{Physical Review B} \textbf{\bibinfo{volume}{46}},
  \bibinfo{pages}{8934} (\bibinfo{year}{1992}).

\bibitem[{\citenamefont{Xian and Merdan}(2014)}]{Xian-JPCS14}
\bibinfo{author}{\bibfnamefont{Y.}~\bibnamefont{Xian}} \bibnamefont{and}
  \bibinfo{author}{\bibfnamefont{M.}~\bibnamefont{Merdan}},
  \bibinfo{journal}{Journal of Physics: Conference Series}
  \textbf{\bibinfo{volume}{529}}, \bibinfo{pages}{012020}
  (\bibinfo{year}{2014}),
  \urlprefix\url{http://stacks.iop.org/1742-6596/529/i=1/a=012020}.

\bibitem[{\citenamefont{Feynman}(1954)}]{feynman-pr54}
\bibinfo{author}{\bibfnamefont{R.~P.} \bibnamefont{Feynman}},
  \bibinfo{journal}{Phys. Rev.} \textbf{\bibinfo{volume}{94}},
  \bibinfo{pages}{262} (\bibinfo{year}{1954}),
  \urlprefix\url{http://link.aps.org/doi/10.1103/PhysRev.94.262}.

\bibitem[{\citenamefont{Wa{\ss}er et~al.}(2016)\citenamefont{Wa{\ss}er, Lee,
  Kihou, Steffens, Schmalzl, Qureshi, and Braden}}]{wasser-16}
\bibinfo{author}{\bibfnamefont{F.}~\bibnamefont{Wa{\ss}er}},
  \bibinfo{author}{\bibfnamefont{C.}~\bibnamefont{Lee}},
  \bibinfo{author}{\bibfnamefont{K.}~\bibnamefont{Kihou}},
  \bibinfo{author}{\bibfnamefont{P.}~\bibnamefont{Steffens}},
  \bibinfo{author}{\bibfnamefont{K.}~\bibnamefont{Schmalzl}},
  \bibinfo{author}{\bibfnamefont{N.}~\bibnamefont{Qureshi}}, \bibnamefont{and}
  \bibinfo{author}{\bibfnamefont{M.}~\bibnamefont{Braden}},
  \bibinfo{journal}{arXiv preprint arXiv:1609.02027}  (\bibinfo{year}{2016}).

\bibitem[{\citenamefont{Zhang et~al.}(2014)\citenamefont{Zhang, Song, Regnault,
  Su, Enderle, Kulda, Tan, Sims, Egami, Si et~al.}}]{zhang-prb14}
\bibinfo{author}{\bibfnamefont{C.}~\bibnamefont{Zhang}},
  \bibinfo{author}{\bibfnamefont{Y.}~\bibnamefont{Song}},
  \bibinfo{author}{\bibfnamefont{L.-P.} \bibnamefont{Regnault}},
  \bibinfo{author}{\bibfnamefont{Y.}~\bibnamefont{Su}},
  \bibinfo{author}{\bibfnamefont{M.}~\bibnamefont{Enderle}},
  \bibinfo{author}{\bibfnamefont{J.}~\bibnamefont{Kulda}},
  \bibinfo{author}{\bibfnamefont{G.}~\bibnamefont{Tan}},
  \bibinfo{author}{\bibfnamefont{Z.~C.} \bibnamefont{Sims}},
  \bibinfo{author}{\bibfnamefont{T.}~\bibnamefont{Egami}},
  \bibinfo{author}{\bibfnamefont{Q.}~\bibnamefont{Si}}, \bibnamefont{et~al.},
  \bibinfo{journal}{Physical Review B} \textbf{\bibinfo{volume}{90}},
  \bibinfo{pages}{140502} (\bibinfo{year}{2014}).

\bibitem[{\citenamefont{Steffens et~al.}(2013)\citenamefont{Steffens, Lee,
  Qureshi, Kihou, Iyo, Eisaki, and Braden}}]{steffens-prl13}
\bibinfo{author}{\bibfnamefont{P.}~\bibnamefont{Steffens}},
  \bibinfo{author}{\bibfnamefont{C.~H.} \bibnamefont{Lee}},
  \bibinfo{author}{\bibfnamefont{N.}~\bibnamefont{Qureshi}},
  \bibinfo{author}{\bibfnamefont{K.}~\bibnamefont{Kihou}},
  \bibinfo{author}{\bibfnamefont{A.}~\bibnamefont{Iyo}},
  \bibinfo{author}{\bibfnamefont{H.}~\bibnamefont{Eisaki}}, \bibnamefont{and}
  \bibinfo{author}{\bibfnamefont{M.}~\bibnamefont{Braden}},
  \bibinfo{journal}{Phys. Rev. Lett.} \textbf{\bibinfo{volume}{110}},
  \bibinfo{pages}{137001} (\bibinfo{year}{2013}),
  \urlprefix\url{http://link.aps.org/doi/10.1103/PhysRevLett.110.137001}.

\end{thebibliography}
\end{document}